\documentclass[draft,jgrga]{agutex}

\usepackage{amsmath}
\usepackage{epsfig}
\usepackage{rotating} 

\usepackage{lineno}

\setkeys{Gin}{draft=false}
\authorrunninghead{WHEELER ET AL.}

\titlerunninghead{ULYSSES ALFV\'{E}N SOLITONS}

\authoraddr{Corresponding author: M. A. Reynolds,
Department of Physical Sciences,
Embry-Riddle Aeronautical University
Daytona Beach, FL 32174, USA.
(anthony.reynolds@erau.edu)}

\begin{document}

%
%

\title{Interpreting Ulysses data using inverse scattering theory: Oblique Alfv\'{e}n waves}
%
%

%
%



\authors{Harry R. Wheeler IV,\altaffilmark{1,2}
M. A. Reynolds,\altaffilmark{1}
and R. L. Hamilton\altaffilmark{3}}

\altaffiltext{1}{Department of Physical Sciences,
Embry-Riddle Aeronautical University,
Daytona Beach, Florida, USA.}

\altaffiltext{2}{Currently at Lockheed Martin Space Systems Company,
Sunnyvale, California, USA.}

\altaffiltext{3}{Department of Math and Applied Sciences,
George Fox University, Newberg, Oregon, USA.}

%
%


\begin{abstract}
Solitary wave structures observed by the Ulysses spacecraft in the solar wind were analyzed using
both inverse scattering theory as well as direct numerical integration of the
derivative nonlinear Schr\"{o}dinger (DNLS) equation.
Several of these structures were found to be consistent with soliton solutions
of the DNLS equation.
Such solitary structures have been commonly observed in the space plasma environment
and may, in fact, be long-lived solitons.
While the generation of these solitons may be due to an instability mechanism,
e.g., the mirror instability,
they may be observable far from the source region due to their coherent nature.
\end{abstract}

%
%

%

\begin{article}

%
%

\section{Introduction}

For several decades, structures in the interplanetary magnetic field,
called
``holes,'' 
``antiholes,'' 
``bumps,'' and
``dips,''
have been observed.
One of the first published observations was of data taken by the magnetometer on the Explorer 43 spacecraft
(also known as Imp I),
where \citet{Turner77} saw localized regions of very low intensity magnetic field (less than 1 nT).
They,
along with \citet{Burlaga78} tendered a theoretical explanation:
these regions were
``stationary nonpropagating equilibrium structures.''
To deduce this,
they used a kinetic boundary layer theory \citep{Lemaire76}
to reproduce the magnetic field profiles that were observed.
Another, more recent, explanation for these structures is that they are
freak waves (or rogue waves).
\citet{Laveder11} has shown that rogue waves appear in randomly driven, dissipative DNLS systems,
but it is not clear how to include these effects in an analysis of the solar wind.
In addition, \citet{Ruderman10} has shown that freak waves can arise in the DNLS context due
to both linear and nonlinear dispersive focusing.

In subsequent observations,
it has become accepted that, because these structures are spatially near regions that are
unstable (or at least near marginal stability) against the mirror instability,
they are mirror mode structures.
As \citet{Winterhalter94} put it,
``there are indications that the holes may have been remnants of mirror-mode structures
created upstream of the points of observation.''
In the two years of Ulysses observations (1991-1992),
\citet{Winterhalter94} found about 30 events per month in which the magnetic field did not change direction across the hole,
a type that they called ``linear holes.''
\citet{Hamilton09} has made the case that these types of holes are well-suited to
analysis using the derivative nonlinear Schr\"{o}dinger (DNLS) equation
and the accompanying inverse scattering transformation (IST).
In addition,
using Wind magnetometer observations,
\citet{Stevens07} also concluded that
``many magnetic holes are remnants of the mirror-mode instability.''

\citet{Rees06} catalogued about 30 events in the Ulysses data set,
i.e., localized increases in the magnetic field strength that had
``the characteristics of a soliton.''
They then compared them with the bright solitons of \citet{Baumgartel99}
which were solutions to the Hall MHD equations, and were similar to the so-called
``one-parameter'' soliton solutions to the DNLS equation.
However, while the properties of the comparison was suggestive,
the agreement between the observations and the analytic solution was not satisfactory.

We will show below that the problem was twofold.
First, for the event that they analyzed in detail (shown in their Fig.~8),
an IST analysis using the transform for the DNLS equation
revealed no eigenvalues, and hence no solitons.
And second,
even if the overall profile shows an increase in the magnetic field strength,
this does not preclude the possibility that the primary soliton embedded in the profile is a dark soliton.

In this paper,
we re-analyze several of the magnetic structures observed by Ulysses and catalogued by
\citet{Rees06},
but we use the suggestion of \citet{Hada93}.
That is, we attack each profile from the perspective
of inverse scattering theory and the DNLS equation.
Each magnetic structure is analyzed in two ways.
First,
the IST technique is used to obtain an analytic
soliton solution that is embedded within the observed magnetic field.
This method searches for eigenvalues of the associated linear eigenvalue problem,
each of which corresponds to either a one-parameter or a two-parameter soliton.
Second,
a direct numerical solution to the DNLS equation is obtained using the observed magnetic field
structure as an initial condition.
A comparison is then made between these two methods.
It is found that in the long-time limit,
after any dispersive component has radiated away,
the numerical solution matches the analytic soliton well.
This result implies that DNLS solitons are the main component of these magnetic structures.

Of course,
it would be more accurate to numerically solve the MHD equations or the two-fluid equations
rather than the DNLS
(to evaluate the time evolution of the initial magnetic field profile).
However,
since Ulysses captures only a snapshot of the magnetic field,
it is more enlightening to compare with the DNLS time evolution,
given that an eigenvalue (or eigenvalues) was found via the IST,
signifying the presence of a soliton.

\section{The derivative nonlinear Schr\"{o}dinger (DNLS) equation}

The DNLS equation is a general nonlinear wave equation that describes
solitons in plasmas under widely varying conditions.
While the standard derivation is for describing quasi-parallel Alfv\'{e}n waves,
in various limits it reduces to the Korteweg-de Vries (KdV) equation
and the modified KdV equation \citep{Kennel88}.
These can be used to describe ion-acoustic solitons, for example \citep{Washimi66}.

A useful form of the DNLS equation is given by \citet{Baumgartel99},
\begin{linenomath}
\begin{equation}
\label{eq:DNLS}
\frac{\partial b}{\partial t} + \alpha \frac{\partial }{\partial x} \left((|b|^2 - |b_0|^2)b\right) +
\frac{i}{2} \frac{\partial^2 b}{\partial x^2} = 0,
\end{equation}
\end{linenomath}
where $b(x,t) = b_y + ib_z$ and
\begin{linenomath}
\begin{equation}
\alpha = \frac{1}{4} \frac{1}{\cos^2\theta-\beta} .
\end{equation}
\end{linenomath}
In this case, the wave propagates in the $x$ direction,
where $x$ is scaled to the ion inertial length $c/\omega_{pi}$,
and Eq.~(\ref{eq:DNLS}) is written in a frame moving in the $x$ direction
with the intermediate Alfv\'{e}n speed $c_A \cos\theta$ relative to the background plasma.
In addition, the magnetic field $b$ has been scaled to the background magnetic field strength $B_0$
away from the soliton.
Also, $\omega_{pi}$ is the ion plasma frequency,
$c$ is the speed of light,
$\beta = c_s^2/c_A^2$ is the plasma ``beta,''
$c_s$ is the plasma sound speed,
$c_A = B_0/\sqrt{\mu_0 \rho}$ is the Alfv\'{e}n speed,
and
$\rho$ is the plasma mass density.
With a suitable redefinition of $x$ and $t$ it is possible to set the
coefficient $\alpha$ equal to unity
(see \citet{Baumgartel99} and \citet{MjolhusHada}),
and we use this redefinition to simplify the numerical calculations.
Other versions of the DNLS equation can be found in \citet{Kennel88} and \citet{Hamilton09}.
The fact that $b_0$ is nonzero reflects the fact that the soliton is not propagating exactly
parallel to the background magnetic field,
and therefore the boundary conditions are ``nonvanishing'' \citep{Kawata78}.

Several analytic solutions to this equation have been found,
and the most well known fall into two classes,
the so-called
one-parameter (either dark or bright)
and two-parameter solitons.
(For explicit expressions for these solutions,
see, for example, \citet{MjolhusHada}
or \citet{Baumgartel99}
or \citet{Hamilton09}.)
These single soliton solutions can also be found using the IST technique,
which uses the fact that some nonlinear wave equations, e.g., Eq.~(\ref{eq:DNLS})
with solution $b(x)$, are associated with Sturm-Liouville-like problems where $b(x)$ takes on the role
of the potential function.
For a given $b(x)$,
the scattering data results in eigenvalues that correspond to soliton solutions of the nonlinear wave equation.
For the DNLS equation,
the IST is a $2 \times 2$ eigenvalue problem,
and the entire procedure is
known as the ``AKNS scheme'' \citep{AblowitzSegur}.
As shown by \citet{Kawata78}, this eigenvalue problem can be written in the form
\begin{linenomath}
\begin{equation}\label{eq:AKNSeigenvalue}
\Phi_x = D \cdot \Phi ,
\end{equation}
\end{linenomath}
where
\begin{linenomath}
\begin{equation}\label{phi}
\Phi (x,t;\lambda)
=
\left( {\begin{array}{cc}
\phi_{11} & \phi_{12} \\
\phi_{21} & \phi_{22}
\end{array} } \right) ,
\end{equation}
\end{linenomath}
and
\begin{linenomath}
\begin{equation}
D = \lambda
\left( {\begin{array}{cc}
-i\lambda & b(x,t) \\
\bar{b}(x,t)& i\lambda
\end{array} } \right)
\mbox{ ,}
\end{equation}
\end{linenomath}
and where $b(x,t)$ is the wave profile of the magnetic field and
$\lambda$ is the eigenvalue.
The one-parameter solitons mentioned above are characterized by real $\lambda$,
and the two-parameter solitons have complex $\lambda$
(the real and imaginary parts of $\lambda$ are the two parameters).
A general magnetic field profile,
such as the Ulysses profiles examined here,
will be composed of solitons (denoted by discrete eigenvalues)
as well as dispersive waves that are part of the continuous spectrum.

To search for eigenvalues of a particular profile $b(x)$,
the relation between the eigenfunctions $\Phi^\pm$
is conventionally given through a scattering matrix
\begin{linenomath}
\begin{equation}
\label{eq;scatrel}
\Phi^-(x,t;\lambda) = \Phi^+(x,t;\lambda) \cdot 
\left( {\begin{array}{cc}
s_{11} & s_{12} \\
s_{21}& s_{22}
\end{array} } \right) ,
\end{equation}
\end{linenomath}
where the elements $s_{ij}(\lambda)$ are functions of the eigenvalue $\lambda$,
and $\Phi^\pm$ are the eigenfunctions of Eq.\ (\ref{eq:AKNSeigenvalue})
defined through their asymptotics as discussed in the Appendix.
Given $b(x)$,
and using the defining asymptotics of $\Phi^\pm$ as boundary conditions,
a numerical solution can be found for $\Phi$ which can then be used to obtain $s_{11}$.
As is standard in scattering matrix theory,
the condition $s_{11}(\lambda) = 0$ determines the set of eigenvalues $\left\{ \lambda_n \right\}$
for which the eigenfunctions are bounded both to the left and right.
The existence of a discrete eigenvalue is essentially the definition that a soliton is embedded in $b(x)$ \citep{DrazinJohnson}.
In addition, dispersive wave solutions to the DNLS
(corresponding to the continuous spectrum of eigenvalues)
can also be present in $b(x)$, 
or they can be the only component.

The numerical technique that we use to solve the DNLS equation is a two-step Lax-Wendroff method for the nonlinear term
with a backward-time, centered-space method for the diffusive term.
The boundary conditions are periodic,
and a large domain is used to ensure that dispersive waves do not 
have time to propagate around
the domain and interfere with the soliton.

\section{Ulysses events}

Several detailed studies of Ulysses observations of magnetic holes have been made
\citep{Tsurutani92,Winterhalter94,Franz00,Rees06}.
Because of the large speed that the spacecraft is moving through the solar wind plasma,
and the small Alfv\'{e}n speed at which the structures are moving through the plasma,
each observation is effectively a snapshot of the magnetic field at one instant of time.

\begin{figure}[p]
\centering
\includegraphics[width=6.5in]{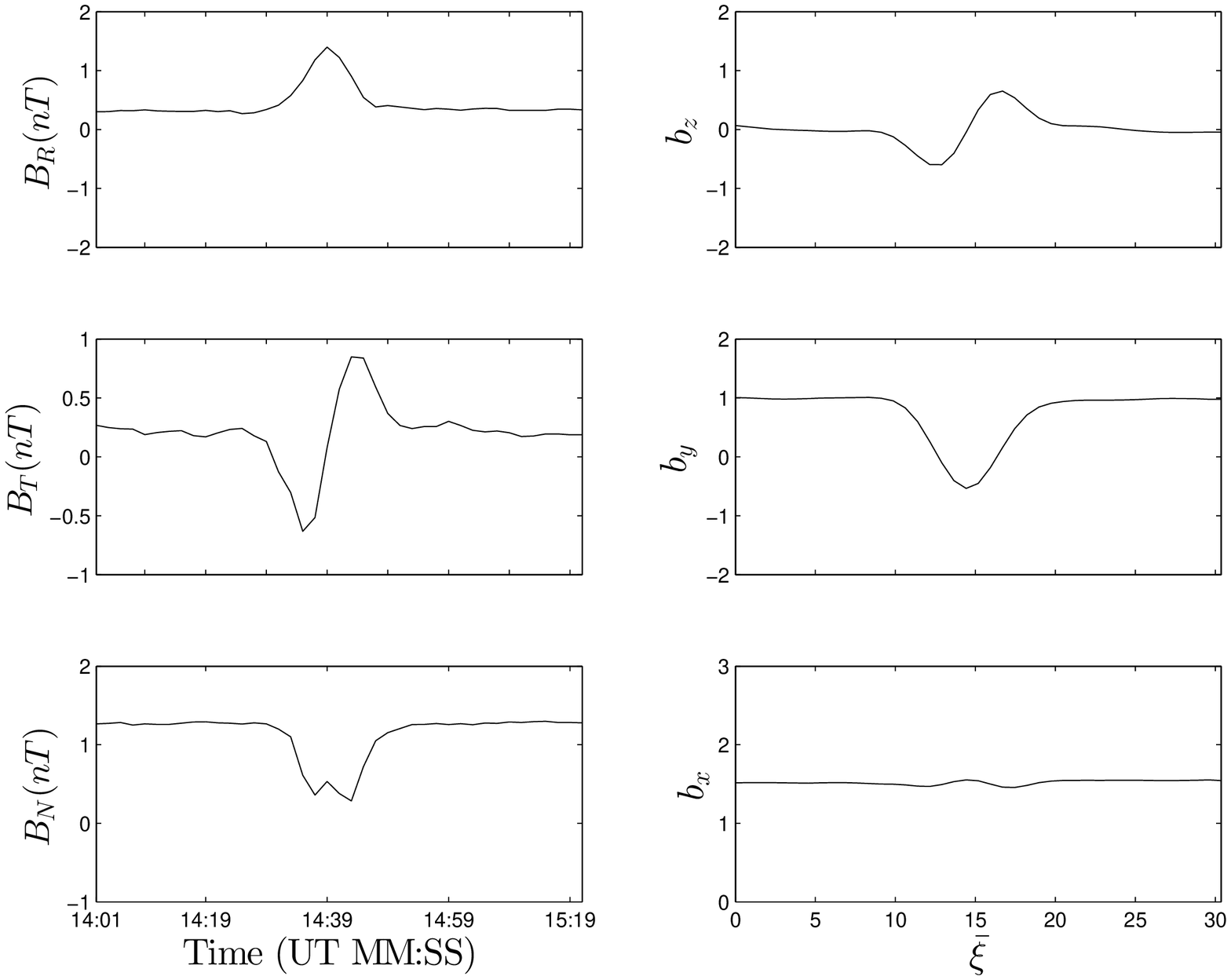}
\caption{Ulysses magnetometer data from 17 July 2002, around 08:15 UT.
(a) Field components as a function of time expressed in RTN coordinates.
(b) Field components as a function of $\xi$, the scaled and rotated $x$,
after the following transformations have been applied:
i) rotation using the minimum variance transformation,
ii) scaling to the asympototic value of $B_x$, and
iii) rotation about the new $x$ direction to eliminate the asymptotic value of $b_z$.
The eigenvalues that resulted from the minimum variance analysis \citep{Sonnerup00} were 33.88, 15.75 and 1,
scaled to the minimum eigenvalue.
The minimum variance direction, $x$, makes an angle of $\theta \approx 32.8^{\circ}$
with respect to the background magnetic field.}
\label{fig:Event2_RawRotated}
\end{figure}

The vector helium magnetometer on Ulysses provided one-second averaged magnetic field measurements,
reported in RTN (Radial-Tangential-Normal) coordinates \citep{Balogh92}.
An example of the raw data, from 17 Jul 2002, is shown in Fig.~\ref{fig:Event2_RawRotated}(a).
The characteristics of the plasma environment during the magnetic field observation,
proton temperature, plasma velocity, and density,
were obtained by the SWOOPS (Solar Wind Observations Over the Poles of the Sun) experiment \citep{Bame92}.
Typical values were $\beta \approx 0.1$, plasma speed $v_{plasma} \approx 500$ km/s,
and ion inertial length $c/\omega_{pi} \approx 10^3$ km.

\citet{Rees06} surveyed the entire Ulysses database looking for soliton-like structures in the magnetic field.
Their criteria were
``short duration increases in the magnetic field magnitude and an associated rotation in the magnetic field direction."
However,
since the magnetic field vector was in the same direction on both sides of the structure,
these are ``linear'' 
in the terminology of \citet{Winterhalter94}.
\citet{Rees06} found 33 events,
and while they appeared to be soliton-like,
the quantitative comparison with the analytic form of a bright, one-parameter soliton
was not ``adequate.''
Here, we re-examine some of the events found by \citet{Rees06} and find that 
the IST results in eigenvalues for many of the events.

The events we have chosen for careful study are shown in Table 1.
The first event listed in Table 1, on 21 Feb 2001, was studied in detail by \citet{Rees06}
(it is shown in their Figure 8).
Because it consisted of a magnetic compression,
they attempted to model it with a bright soliton solution of the DNLS equation \citep{Baumgartel99}.
However,
as can be seen from Fig.~8 of \citet{Rees06},
the fit was not very good.
Our explanation for this poor agreement is that
there is no DNLS soliton embedded in the profile, which is verified by the lack of a discrete eigenvalue found by our
scattering analysis.
As mentioned above, 
a discrete eigenvalue determines the presence of a soliton,
and the lack of any eigenvalue implies that there are only
dispersive wave solutions to the DNLS in the magnetic field profile.
The subsequent evolution of that magnetic field profile, as determined by a numerical solution of the DNLS,
shows that the initial pulse collapses into a wave train.
Another reason for the lack of a DNLS soliton is that this wave was propagating almost perpendicular
to the background magnetic field.
This is a regime where the DNLS equation is thought not to be applicable for low $\beta$ plasmas.
See, for example, the discussions in \citet{Baumgartel99}, \citet{Buti01}, and \citet{Ruderman02}.

\begin{figure}[p]
\centering
    \includegraphics[width=4.5in]{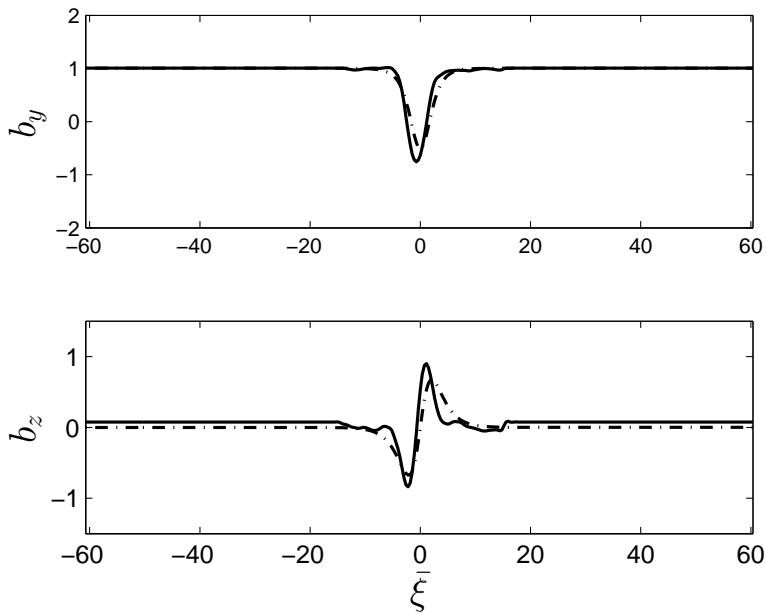}
    \caption{Solid lines: components of $b$ from Fig.~\ref{fig:Event2_RawRotated}(b).
    Dashed lines: analytic dark soliton solution of the DNLS (eigenvalue of $\lambda \approx 0.223$) at $t=0$.
    Note that
    the components of $b$ at $t=0$ have been padded and the domain increased
    so that the dispersive waves that are also present in the profile do not have time to propagate
    around the domain and interfere with the soliton.}
\label{fig:Event2_t0}
\end{figure}

\begin{figure}[p]
\centering
    \includegraphics[width=4.5in]{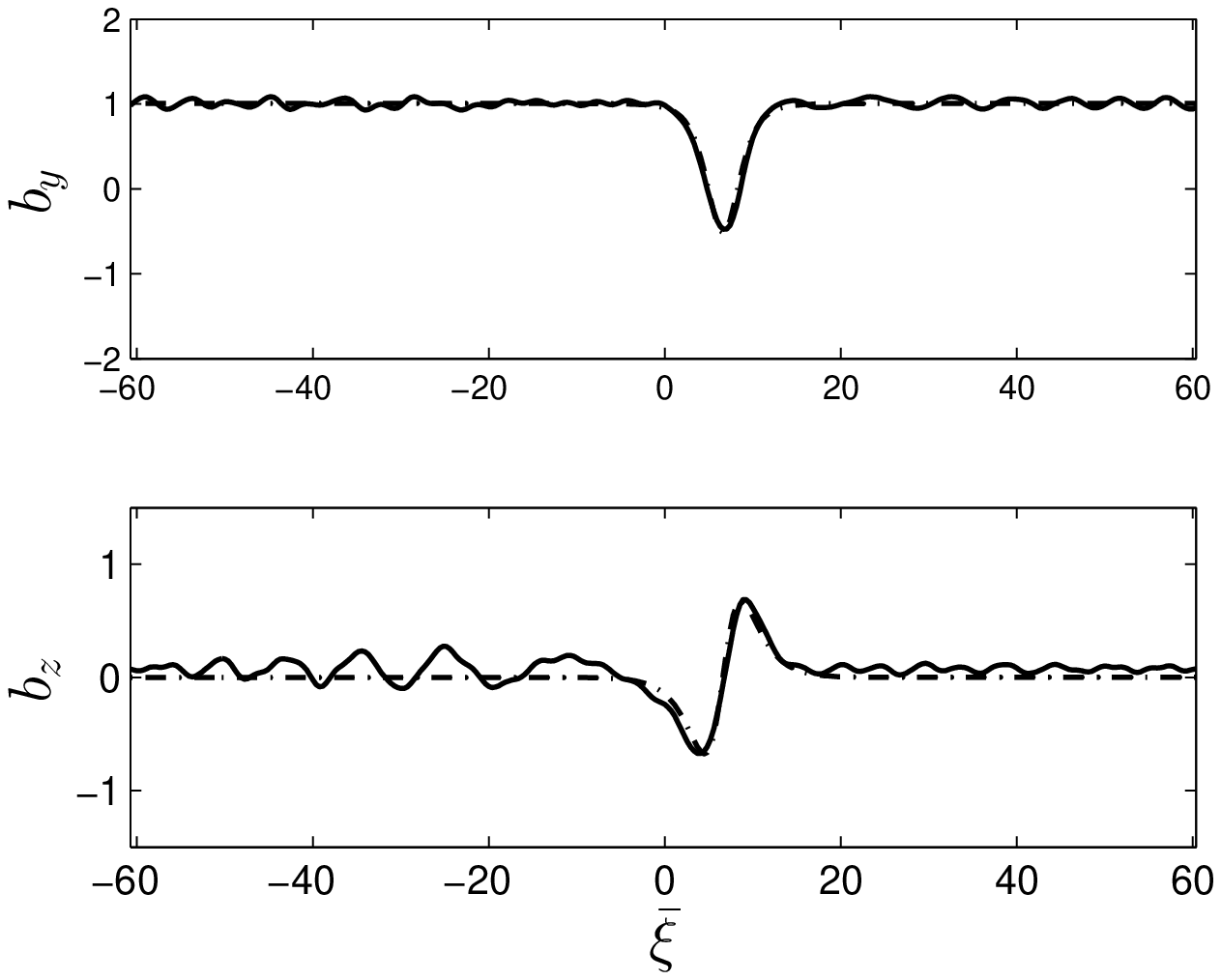}
    \caption{Same as Fig.~\ref{fig:Event2_t0}, but at $t = 55.2 \, \Omega_{p}^{-1}$.
    The observed magnetic field (solid lines) have been evolved using the numerical solution of the DNLS,
    while the analytic solution (dashed lines) has the time dependence included.
    The analytic solution travels at a speed between the Alfv\'{e}n speed and the magnetosonic speed.
    The particular time was chosen to allow the dispersive waves enough time to propagate away from the soliton, 
    but not enough time to wrap around the domain.}
\label{fig:Event2_t15}
\end{figure}

The second event listed in Table 1, on 17 July 2002, is shown in Figs.~\ref{fig:Event2_RawRotated}--\ref{fig:Event2_t15}.
Figure~\ref{fig:Event2_RawRotated} (a) shows the raw magnetometer data,
while \ref{fig:Event2_RawRotated} (b) shows the rotated, smoothed, and scaled field components.
A minimum variance analysis \citep{Sonnerup00} was performed to determine the orientation of the structure,
and the $x$-direction was scaled in order to use Eq.~(\ref{eq:DNLS}) with $\alpha=1$.
Finally,
the profile was smoothed using exponential smoothing in order to simplify the root-finding algorithm that 
searched for eigenvalues.

Figure~\ref{fig:Event2_t0} shows the magnetic field structure from Fig.~\ref{fig:Event2_RawRotated}(b),
which is the initial condition,
along with a large-amplitude one-parameter dark soliton with eigenvalue $\lambda_1=0.223$.
There was a second real eigenvalue found by the scattering analysis,
but this had a very small amplitude since its eigenvalue was close to unity, $\lambda_2 = 0.994$,
and in addition, there were dispersive waves.
That the $\lambda_1=0.223$ soliton is the most significant can be seen from the energy
in the magnetic field 
\begin{linenomath}
\begin{equation}
\label{eq;energy}
E = \int_{-\infty}^\infty \left( |b|^2 - b_0^2 \right) dx .
\end{equation}
\end{linenomath}
This integral can be analytically evaluated for solitons \citep{Sanchez10}
and predicts that the total energy of the two dark solitons is $E_1=-5.38$ and $E_2=-0.438$,
respectively,
and this was confirmed numerically.
In addition,
the energy of the dispersive waves, which can only be found numerically, is $E_{disp} = 3.28$.
This shows that the second soliton has minimal energy compared with the dispersive waves,
and once they propagate away
the solitary structure is effectively a single soliton.

This can be seen in Fig.\ \ref{fig:Event2_t15},
which shows the observed magnetic field and the DNLS soliton at time $55.2 \, \Omega_p^{-1}$ later,
where $\Omega_p$ is the proton cyclotron frequency.
The Ulysses magnetic field has been propagated forward in time with a numerical solution of the DNLS,
and compared with the analytic time-dependent dark soliton solution.
Note that the soliton matches the observed magnetic field much more closely at this later time,
because the dispersive wave component has propagated away toward the edges of the simulated domain.
Also,
this analysis answers the question of why a dark soliton represents a magnetic increase so well.
The reason is that (in this case)
the compressive magnetic field profile consists of a large-amplitude dark soliton,
a small-amplitude dark soliton,
and a dispersive wave component
with significant energy, as shown above.
The stability of dark solitons allow them to remain embedded in the magnetic field,
while any bright solitons decay away.

\section{Discussion}

It is instructive to note that all of the events in Table 1
are characterized by multiple eigenvalues.
This simply means that the wave form of the magnetic field is comprised of 
multiple solitons.
Of course, the combination of solitons 
is not linear, but nonlinear,
and general methods to obtain so-called ``$N$-soliton solutions'' can be found in
\citet{AblowitzSegur} and \citet{DrazinJohnson}, for example.
In certain cases, such as the one examined in detail in Figs. 1-3,
only one soliton contributes significantly to the magnetic field profile.
In general, however, the nonlinear $N$-soliton solution must be compared with the observed profile.
It is important to note also that the existence of a discrete eigenvalue for an arbitrary profile, $b(x,t)$, 
does not necessarily mean that the profile is a soliton. 
It simply means that there is a soliton embedded in the profile, possibly along with
another soliton (or solitons) or a dispersive wave component, or both.

As long as the underlying nonlinear dynamics are captured by the DNLS, then solitons are the most fundamental way of representing the wave dynamics.
In addition,
if the plasma conditions are consistent with dissipation or nonlinear Landau damping or even random density fluctuations, 
then it is plausible that the multiple solitons have a common origin.
That is,
the damping of a bright one-parameter soliton will inevitably give rise to the formation of a two-parameter
soliton which will subsequently damp away itself and form a train of dark one-parameter solitons in the 
process \citep{Sanchez10}.
This naturally leads to the question of how the solitons that we have identified,
moving with different speeds,
are able to arrive simultaneously from a distant source location.
Of course, 
within the context of the pure DNLS equation,
they cannot.
However, 
the physical effects just listed are not included in the DNLS,
and
it is a challenge for future work
to determine if a combination of these effects could maintain a 
sufficiently coherent wave packet comprised of mostly dark solitons.

As stated by \citet{Hada93},
this method of analysis is ``more efficient than the Fourier transform'' when analyzing nonlinear waves.
They go on to point out that there are several other advantages to the inverse scattering analysis,
including the fact that both the past history and future evolution of the nonlinear waves is contained
in the scattering data.
With a single spacecraft, unfortunately, it is at best difficult to prove these connections.
However, both the current Cluster II set of satellites and the future Magnetospheric Multiscale mission
provide opportunities for such analysis.
In fact,
\citet{Stasiewicz03} showed that there was good agreement between Cluster observations
in the magnetopause boundary layer and magnetosonic solitons,
and
\citet{Trines07} saw good agreement between simulations and Cluster observations at the magnetopause
of solitary electrostatic structures.
However, we have shown here that a deeper understanding of nonlinear waves 
can be obtained using the 
inverse scattering analysis to obtain eigenvalues,
in addition to a direct comparison between observations and theory.


%
%
%

\appendix

\section*{Appendix: Inverse Scattering Transform of the DNLS}

For the case of nonvanishing boundary conditions
(i.e., when $b \rightarrow b_0$ as $x\rightarrow \pm\infty$),
\citet{Kawata78} showed that 
the linear eigenvalue equation in Eq.\ (\ref{eq:AKNSeigenvalue}), 
$\Phi_x = D \cdot \Phi$,
has the asymptotic eigenfunctions
\begin{linenomath}
\begin{equation} \label{BCs}
\Phi^{\pm}(x,t; \lambda) \rightarrow T(\lambda,\zeta)J(\Lambda x) \mbox{ when } x \rightarrow \pm \infty ,
\end{equation}
\end{linenomath}
where
\begin{linenomath}
\begin{equation}
T(\lambda,\zeta)
 =
\left( {\begin{array}{cc}
-ib_0 & \lambda - \zeta \\
\lambda - \zeta& ib_0
\end{array} } \right)
 \mbox{ ,}
\end{equation}
\end{linenomath}
and
\begin{linenomath}
\begin{equation}
J(\Lambda x)
 =
\left( {\begin{array}{cc}
e^{-i\Lambda x} & 0 \\
0 & e^{i\Lambda x}
\end{array} } \right) ,
\end{equation}
\end{linenomath}
and where
$\Lambda = \lambda\zeta$
and 
$\zeta = \sqrt{\lambda^2 - b_0^2}$.
Since $\zeta$ and $\Lambda$ are complex,
the root chosen to express $\Phi^\pm$ is that which results in a bound eigenfunction,
i.e., $\operatorname{Im} (\Lambda) >0$.
Then,
as is mentioned in the text,
the scattering matrix $\mathsf{S}$ is defined by
$\Phi^- = \Phi^+ \cdot \mathsf{S}$,
and the requirement that $s_{11}=0$
results in the set of eigenvalues for which $\Phi$ is bounded.

%
%
%
%

\begin{acknowledgments}
The authors would like to thank the Ulysses Final Archive
(\texttt{http://ufa.esac.esa.int/ufa/})
for the preservation of the Ulysses data set,
and for the free distribution of that data.
In addition, the authors would like to thank the referees for important comments
that improved the paper.
H.R.W. would like to thank Andrei Ludu for useful discussions.
\end{acknowledgments}

\end{article}
%
%
%
%
%
%

\newpage

%
%


\newpage

\begin{table}[h]
\label{table:EventProperties}
\caption{Properties of selected events.
Columns are
date and time (UT),
$V$ is the s/c velocity relative to the surrounding plasma,
$\beta$ is the plasma beta,
$\theta$ is the angle between the background magnetic field and the layer normal $\hat{k}$,
and $v_A$ is the Alfv\'{e}n speed of the surrounding plasma.
The last two columns list the number of one and two-parameter eigenvalues found for each profile.
}
\centering
\begin{tabular}{llcccccc}
\hline
 Date & Time & $V$ & $\beta$ & $\theta$ & $v_A$ & one- & two- \\
 & (UT) & ($10^5$ m/s) & & & (km/s) & parameter & parameter \\
\hline
 2001 Feb 21 & 03:35 & 5.19 & 0.498 & 93.1$^\circ$ & 38.99 & &  \\
 2002 Jul 17 & 08:15 & 4.96 & 0.085 & 32.8$^\circ$ & 98.28 & 2 &  \\
 1991 Jul 02 & 21:10 & 3.85 & 0.153 & 31.3$^\circ$ & 88.86 & 1 & 1 \\
 1991 Jan 05 & 04:34 & 2.68 & 0.058 & 26.6$^\circ$ & 103.8 & 3 & 1 \\
 1991 Apr 01 & 04:57 & 5.61 & 0.123 & 15.1$^\circ$ & 83.76 & 5 & 3 \\
 1992 Jan 14 & 19:56 & 5.15 & 0.156 & 178.2$^\circ$ & 88.39 & 10 & 10 \\
 2001 Feb 21 & 06:13 & 5.07 & 0.444 & 35.5$^\circ$ & 38.05 & & 7 \\
 2001 Feb 21 & 09:23 & 4.85 & 0.440 & 6.87$^\circ$ & 34.77 & 10 & 1 \\
 2001 Feb 21 & 09:28 & 4.79 & 0.437 & 16.54$^\circ$ & 35.88 & 2 &  \\
\hline
\end{tabular}
\end{table}


\begin{thebibliography}{}

\providecommand{\natexlab}[1]{#1}
\expandafter\ifx\csname urlstyle\endcsname\relax
  \providecommand{\doi}[1]{doi:\discretionary{}{}{}#1}\else
  \providecommand{\doi}{doi:\discretionary{}{}{}\begingroup
  \urlstyle{rm}\Url}\fi
%

\bibitem[{\textit{Ablowitz and Segur}(1981)}]{AblowitzSegur}
Ablowitz, M. J., and H. Segur (1981),
\textit{Solitons and the Inverse Scattering Transform}, SIAM, Philadelphia.

\bibitem[{\textit{Balogh et~al.}(1992)}]{Balogh92}
Balogh, A., T. J. Beek, R. J. Forsyth, P. C. Hedgecock, R. J. Marquedant, E. J. Smith, D. J. Southwood, and B. T. Tsurutani (1992),
The magnetic field investigation on the ULYSSES mission: Instrumentation and preliminary scientific results,
{\it Astron.\ Astrophys.\ Suppl.\ Ser.}, \textit{92}(2), 221-236.

\bibitem[{\textit{Bame et~al.}(1992)}]{Bame92}
Bame, S. J.,  D. J. McComas, B. L. Barraclough, J. L. Phillips,  K. J. Sofaly, and  J. C. Chavez (1992),
The Ulysses solar wind plasma experiment,
{\it Astron.\ Astrophys.\ Suppl.\ Ser.}, \textit{92}, 237-265.

\bibitem[{\textit{Baumg\"artel}(1999)}]{Baumgartel99}
Baumg\"artel, K. (1999),
Soliton approach to magnetic holes,
\textit{J. Geophys.\ Res.}, \textit{104}, 28295-28308.

\bibitem[{\textit{Burlaga and Lemaire}(1978)}]{Burlaga78}
Burlaga, L. F., and J. F. Lemaire (1978),
Interplanetary magnetic holes: Theory,
\textit{J. Geophys.\ Res.}, \textit{83}, 5157-5160.

\bibitem[{\textit{Buti et~al.}(2001)}]{Buti01}
Buti, B., B. T. Tsurutani, M. Neugebauer, and B. E. Goldstein (2001),
Generation mechanism for magnetic holes in the solar wind,
\textit{Geophys.\ Res. Lett.}, \textit{28}, 1355-1358.

\bibitem[{\textit{Drazin and Johnson}(1989)}]{DrazinJohnson}
Drazin, P. G., and R. S. Johnson (1989),
\textit{Solitons: An Introduction}, Cambridge University Press, Cambridge.

\bibitem[{\textit{Franz et~al.}(2000)}]{Franz00}
Franz, M., D. Burgess, and T. S. Horbury (2000),
Magnetic field depressions in the solar wind,
\textit{J. Geophys.\ Res.}, \textit{105}, 12725-12732.

\bibitem[{\textit{Hada et~al.}(1993)\textit{Hada, Hamilton, and Kennel}}]{Hada93}
Hada, T., R. L. Hamilton, and C. F. Kennel (1993),
The soliton transform and a possible application to nonlinear Alfv\'{e}n waves in space,
\textit{Geophys.\ Res. Lett.}, \textit{20}, 779-782.

\bibitem[{\textit{Hamilton et~al.}(2009)\textit{Hamilton, Peterson, and Libby}}]{Hamilton09}
Hamilton, R. L., D. A. Peterson, and S. M. Libby (2009),
Magnetic hole formation from the perspective of inverse scattering theory,
\textit{J. Geophys.\ Res.}, \textit{114}, A03104, \doi{10.1029/2008JA013582}.

\bibitem[{\textit{Kawata and Inoue}(1978)}]{Kawata78}
Kawata, T., and H. Inoue (1978),
Exact solution of the derivative nonlinear Schr\"{o}dinger equation under the nonvanishing conditions,
\textit{J. Phys. Soc. Jpn.}, \textit{44}(6), 1968-1976.

\bibitem[{\textit{Kennel et~al.}(1988)\textit{Kennel, Buti, Hada, and Pellat}}]{Kennel88}
Kennel, C. F., B. Buti, T. Hada, and R. Pellat (1988),
Nonlinear, dispersive, elliptically polarized Alfv\'{e}n waves,
\textit{Phys. Fluids}, \textit{31}, 1949-1961.

\bibitem[{\textit{Laveder et~al.}(2011)}]{Laveder11}
Laveder, D., T. Passot, P. L. Sulem, and G. S\'{a}nchez-Arriaga (2011),
Rogue waves in Alfv\'{e}nic turbulence,
\textit{Phys. Lett. A}, \textit{375}, 3997-4002.

\bibitem[{\textit{Lemaire and Burlaga}(1976)}]{Lemaire76}
Lemaire, J. F., and L. F. Burlaga (1976),
Diamagnetic boundary layers: A kinetic theory,
\textit{Astrophys. Space Sci.}, \textit{45}, 303-325.

\bibitem[{\textit{Mj\o lhus and Hada}(1997)}]{MjolhusHada}
Mj\o lhus, E., and T. Hada (1997), Soliton theory of quasi-parallel MHD waves,
in \textit{Nonlinear waves and Chaos in Space Plasmas},
edited by T. Hada and H. Matsumoto, pp. 121-169, Terra Sci., Tokyo.

\bibitem[{\textit{Rees et~al.}(2006)\textit{Rees, Balogh, and Horbury}}]{Rees06}
Rees, A., A. Balogh, and T. S. Horbury (2006),
Small-scale solitary wave pulses observed by the Ulysses magnetic field experiment,
\textit{J. Geophys.\ Res.}, \textit{111}, A10106, \doi{10.1029/2005JA011555}.

\bibitem[{\textit{Ruderman}(2002)}]{Ruderman02}
Ruderman, M. S. (2002),
DNLS equation for large-amplitude solitons propagating in an arbitrary direction in a high-$\beta$ Hall plasma,
\textit{J. Plasma Phys.}, \textit{67}, 271-276.

\bibitem[{\textit{Ruderman}(2010)}]{Ruderman10}
Ruderman, M. S. (2010),
Freak waves in laboratory and space plasmas,
\textit{Eur. Phys. J. -- Spec. Top.}, \textit{185}, 57-66.

\bibitem[{\textit{S\'{a}nchez-Arriaga}(2010)}]{Sanchez10}
S\'{a}nchez-Arriaga, G. (2010),
Alfven soliton and multisoliton dynamics perturbed by nonlinear Landau damping,
\textit{Phys.\ Plasmas}, \textit{17}, 082313.

\bibitem[{\textit{Sonnerup and Scheible}(2000)}]{Sonnerup00}
Sonnerup, B.U., and M. Scheible (2000),
Minimum and Maximum Variance Analysis,
in {\it Analysis Methods for Multi-Spacecraft Data}, edited by G. Paschmann and P. Daly, pp.\ 185-220,
Inter. Space Science Inst., Noordwijk.

\bibitem[{\textit{Stasiewicz et~al.}(2003)}]{Stasiewicz03}
Stasiewicz, K., P. K. Shukla, G. Gustafsson, S. Buchert, B. Lavraud, B. Thid\'{e}, and Z. Klos (2003),
Slow magnetosonic solitons detected by the Cluster spacecraft,
\textit{Phys.\ Rev. Lett.}, \textit{90}, 085002, \doi{10.1103/PhysRevLett.90.085002}.

\bibitem[{\textit{Stevens and Kasper}(2007)}]{Stevens07}
Stevens, M. L., and J. C. Kasper (2007),
A scale-free analysis of magnetic holes at 1 AU,
\textit{J. Geophys.\ Res.}, \textit{112}, A05109, \doi{10.1029/2006JA012116}.

\bibitem[{\textit{Trines et~al.}(2007)}]{Trines07}
Trines, R., R. Bingham, M. W. Dunlop, A. Vaivads, J. A. Davies, J. T. Mendon\c{c}a, L. O. Silva, and P. K. Shukla (2007),
Spontaneous generation of self-organized solitary wave structures at Earth's magnetopause,
\textit{Phys.\ Rev. Lett.}, \textit{9}, 205006, \doi{10.1103/PhysRevLett.99.205006}.

\bibitem[{\textit{Tsurutani et~al.}(1992)}]{Tsurutani92}
Tsurutani, B. T., D. J. Southwood, E. J. Smith, and A. Balogh (1992),
Nonlinear magnetosonic waves and mirror mode structures in the March 1991 Ulysses interplanetary event,
\textit{Geophys.\ Res. Lett.}, \textit{19}, 1267-1270.

\bibitem[{\textit{Turner et~al.}(1977)}]{Turner77}
Turner, J. M., L. F. Burlaga, N. F. Ness, and J. F. Lemaire (1977),
Magnetic holes in the solar wind,
\textit{J. Geophys.\ Res.}, \textit{82}, 1921-1924.

\bibitem[{\textit{Washimi and Taniuti}(1966)}]{Washimi66}
Washimi, Hariuchi, and Tosiya Taniuti (1966),
Propagation of Ion-Acoustic Solitary Waves of Small Amplitude,
\textit{Phys.\ Rev.\ Lett.}, \textit{17}, 996-–998.

\bibitem[{\textit{Winterhalter et~al.}(1994)}]{Winterhalter94}
Winterhalter, D., M. Neugebauer, B. E. Goldstein, E. J. Smith, S. J. Bame, and A. Balogh (1994),
Ulysses field and plasma observations of magnetic holes in the solar wind and their relation to mirror-mode structures,
\textit{J. Geophys.\ Res.}, \textit{99}, 23371-23381.

-----------------------------------------------------------------------


\end{thebibliography}
\end{document}